\documentclass[manuscript,nonacm]{acmart}

\AtBeginDocument{%
  }

\setcopyright{acmlicensed}
\copyrightyear{2024}
\acmYear{2024}
\acmDOI{XXXXXXX.XXXXXXX}

\acmConference[CHI `24]{}{May 11--16,
  2024}{Honolulu, HI}
\acmISBN{978-1-4503-XXXX-X/18/06}

\acmSubmissionID{2857}

\usepackage{tabularx}
\usepackage{tabulary}
\usepackage{capt-of}

\usepackage{csquotes}

\usepackage{color, colortbl}

\usepackage{makecell}

\usepackage{booktabs,siunitx}

\usepackage{lipsum}

\usepackage{enumitem}

\usepackage{xspace}

\usepackage[normalem]{ulem}
\definecolor{mred}{rgb}{.80,.12,.30}
\definecolor{MRED}{rgb}{.80,.12,.30}
\definecolor{grey}{rgb}{0.5,0.5,0.5}
\definecolor{purple}{rgb}{.75,0,.85}
\definecolor{pistachio}{rgb}{0.58, 0.77, 0.45}
\definecolor{palesilver}{rgb}{0.9, 0.9, 0.9}

\newif\ifnotes
\notestrue

\let\origcite\cite
\renewcommand{\cite}[1]{\ifnotes\mbox{\origcite{#1}}\else \origcite{#1}\fi}

\begin{document}

\title[Don't Translate, Advocate: Using Large Language Models as Devil's Advocates for AI Explanations]{Don't Just Translate, Agitate: Using Large Language Models as Devil's Advocates for AI Explanations}


\author{Ashley Suh}
\email{ashley.suh@mit.ll.edu}

\author{Kenneth Alperin}
\email{steven.gomez@ll.mit.edu}

\author{Harry Li}
\email{harry.li@mit.ll.edu}

\author{Steven R Gomez}
\email{steven.gomez@ll.mit.edu}

\affiliation{%
  \institution{MIT Lincoln Laboratory}
  \streetaddress{244 Wood Street}
  \city{Lexington}
  \state{MA}
  \country{USA}
  \postcode{02421}
}

\renewcommand{\shortauthors}{Suh et al.}

\begin{abstract}
This position paper highlights a growing trend in Explainable AI (XAI) research where LLMs are used to \textit{translate} outputs from explainability techniques, like feature-attribution weights, into a natural language explanation. While this approach may improve accessibility or readability for users, recent findings suggest that translating into human-like explanations does not necessarily enhance user understanding and may instead lead to overreliance on AI systems. When LLMs summarize XAI outputs without surfacing model limitations, uncertainties, or inconsistencies, they risk reinforcing the illusion of interpretability rather than fostering meaningful transparency.
We argue that---instead of merely translating XAI outputs---LLMs should serve as constructive agitators, or \textit{devil's advocates}, whose role is to actively interrogate AI explanations by presenting alternative interpretations, potential biases, training data limitations, and cases where the model's reasoning may break down. In this role, LLMs can facilitate users in engaging critically with AI systems and generated explanations, with the potential to reduce overreliance caused by misinterpreted or specious explanations.
\end{abstract}



\keywords{Large language models, explainable AI, trustworthy AI, interpretability. \\
\textbf{Presented at the Human-centered Explainable AI Workshop (HCXAI) @ CHI 2025.} DOI: \href{https://doi.org/10.5281/zenodo.15170455}{10.5281/zenodo.15170455}
}

\maketitle

\section{Introduction}
\label{sec:intro}
The emergence of Large Language Models (LLMs) and multimodal LLMs~\cite{zhang2024mm} as versatile reasoning tools has driven a recent trend of integrating them into Explainable AI (XAI) workflows. The prevailing approach is to feed XAI outputs (e.g., encodings of feature attributions generated from frameworks like SHAP~\cite{lundberg2017unified} or LIME~\cite{ribeiro2016should}) into these generative models, and then prompt the models to create more natural-sounding translations of the explanations~\cite{zytek2024llms, zytek2024explingo, khediri2024enhancing, crisan2024exploring, fredes2024using, giorgi2024natural, he2025conversational}. There are several motivations for this approach, beyond accessibility:
\begin{itemize}[topsep=0pt, partopsep=0pt,itemsep=0pt,parsep=0pt]
    \item \textit{Transforming the user task from synthesis and agreement, to just agreement.} Many people want to interpret explanations as natural-language \textit{narratives} for improved easy of interpretability~\cite{zytek2024explingo}, but popular explainer methods like SHAP do not output explanations that can be easily transformed into useful narratives. Users must first interpret the raw explanation and then decide whether it supports the model’s decision. 
    \item \textit{``When you've got it, flaunt it."} Generative AI tools like ChatGPT, Copilot, etc., are becoming commonplace tools in work environments where XAI is readily-deployable today for ML developers. At a glance, translating structured explanation outputs into text narratives is the kind of thing GenAI \textit{should} be good at---relying on massive language training to form intelligible, context-bearing sentences around feature information.
\end{itemize}
While these motivations are reasonable, we argue that LLMs could exacerbate---rather than help---known risks with XAI if the LLM is not integrated in careful, strategic ways.

In this workshop paper, we examine how LLMs are currently being integrated for XAI and outline an alternative path forward. In Sec.~\ref{sec:today}, we highlight from the literature where cracks emerge in LLM-based explanation translation and dialogue support for counterfactuals. 
In Sec.~\ref{sec:prompting}, we introduce a contrary approach to these previous works: rather than instructing LLMs to produce agreeable explanations, we provide jumping-off points for few-shot training an LLM-based \textit{devil's advocate}~\cite{chiang2024enhancing} specifically for XAI---one that actively challenges, contextualizes, and problematizes model outputs.
Finally, in Sec.~\ref{sec:tomorrow} we discuss future opportunities for LLM-based approaches to better align with the goals of the underlying decision-support mission and foster critical engagement rather than passive agreement.

\section{LLMs for XAI, Today}
\label{sec:today}
While XAI has seen success, especially ``left of deployment" (e.g., helping ML developers discover and correct spurious model behaviors~\cite{lundberg2017unified, ribeiro2016should}), there remains challenges for users who rely on AI explanations for decision support. People using AI systems as tools have limited time and attention, and they often prioritize their decision tasks over sanity-checking the AI. Moreover, they bring with them a panoply of cognitive biases related to information processing and sensemaking, and an AI-generated explanation is not immune from those biases. 
Recent work has focused on using LLMs with the goal of making them \textit{less prone to interpretation errors} than the raw XAI output~\cite{zhao2023survey}---we discuss further below.
\\ \\
\noindent\textbf{Producing Explanation Narratives and Summaries}.
A typical approach is to use LLMs to create natural language explanations from structured feature-attribution explanations in order to make them more accessible and easier to understand. Various pipelines have been proposed to ensure that these natural language explanations are intelligible. For example, Zytek et al.~quantitatively demonstrated LLMs can construct good quality explanation narratives from SHAP feature values~\cite{lundberg2017unified}, and showed that users overall preferred LLM-based narrative explanations over plots of SHAP values in a comparative study (N=20) due to their ease of understanding and informativeness~\cite{zytek2024llms}. 
Their resultant system Explingo~\cite{zytek2024explingo} uses a pipeline of \textit{narrator} and \textit{grader} LLM components they developed to produce and score candidate explanation narratives from the SHAP features values produced alongside the classifier decision. 

A key finding was that having few-shot exemplar narratives that map between the structured SHAP values and a narrative---both hand-written and bootstrapped ones (e.g., produced automatically to fulfill baseline scoring criteria from the \textit{grader} component)---were important to ensure that new LLM narratives were sensible to human evaluators. In other words, the LLM alone could not translate between the SHAP vectors to natural language while maintaining the criteria of accuracy, conciseness, completeness, and fluency. Consequently, few-shot exemplars may be difficult to write a priori, and their effectiveness or completeness as exemplars might depend on the decision domain.
\\ \\
\noindent\textbf{AI Explanations in High-Stakes Decision Domains}. Similar narrative-generation approaches have been applied in complex decision domains like classifiers for cybersecurity (e.g., network intrusion detection~\cite{khediri2024enhancing}). In these scenarios, a user's assessment about an AI output may greatly impact their downstream actions, so explanations should help a user reject the output when the model is wrong. But studies often lack convincing evaluations to demonstrate that narratives help users detect AI errors better than the AI outputs alone or `raw' explainer-generated explanations. 
\\ \\
\noindent\textbf{Presenting Explanations of Counterfactual Explanations}. LLMs have also been used to summarize explanations focused on helping users interpret model behaviors through \textit{counterfactual} examples. Crisan et al.~\cite{crisan2024exploring} conducted a small study (N=10) to explore how different counterfactual explanations~\cite{wachter2017counterfactual} (which were presented with a visualization of LIME~\cite{ribeiro2016should} and an LLM-generated summary) affected users' perceptions of the underlying model's correctness, fairness, and trustworthiness.
Overall, the study demonstrated that LLM-presented counterfactual example (CFE) outputs with LIME do not inherently improve understanding or trust calibration. Participants often struggled to make sense of LLM-explained LIME feature attributions, finding that numerical importance scores lacked meaningful context and that LLM-generated summaries made explanations sound more plausible rather than more transparent. Some users accepted the explanations at face value, while others dismissed them when they clashed with their mental model of the data, revealing a tendency toward confirmation bias rather than critical engagement.
\\ \\
\noindent\textbf{Conversational LLM-Based XAI Assistants}.
Interestingly, large language models as XAI assistants can now support freeform questions from the user rather than generating a `one size fits all' explanation. This is a compelling idea, but so far research shows that overreliance on AI~\cite{buccinca2021trust} is not necessarily solved by using an LLM. Recently, He et al.~\cite{he2025conversational} studied the effects of a conversational XAI assistant, in which users could engage in a back-and-forth with an LLM for their explainability questions. The authors conducted a large-scale empirical study (N=306) in a loan approval task, comparing this LLM-based conversational XAI assistant to a static XAI dashboard as well as a rule-based conversational XAI assistant. Four XAI methods were employed (SHAP, PDP~\cite{friedman2001greedy}, Mace~\cite{yang2022mace}, the What-If Tool~\cite{whatif}, and a Decision Tree~\cite{dt}), enabling the assistant to answer a variety of questions.
Unsurprisingly, the LLM-based XAI assistant produced natural and persuasive explanations, which resulted in higher user engagement and perceived trust in the AI system. However, the authors found that these highly plausible LLM-generated responses led to users misplacing their trust in the AI system, even when its predictions were incorrect---aligning with previous findings in which users overrelied on incorrect AI suggestions, even when provided with explanations~\cite{buccinca2021trust}.  
\\ \\
\textbf{Takeaways}. While these findings reveal possible failure modes for LLMs in XAI, we argue in Sec.~\ref{sec:prompting} that reframing the role of an LLM in XAI into one that challenges the structured explanation outputs from XAI (and where users are encouraged to interpret the LLM's outputs with healthy skepticism) can create an upside from the general limitations of LLMs, like the tendency to smooth over model inconsistencies persuasively rather than surface them~\cite{he2025conversational}.


\section{Nudging a Helpful \textit{Assistant} Into a Helpful \textit{Adversary}}
\label{sec:prompting}
The explainability pitfalls mentioned in Sec.~\ref{sec:today} risk being worsened with LLMs, as they are often instructed to explain XAI outputs as ``simply'' and ``naturally'' as possible to non-experts~\cite{zytek2024llms}. Consequently, users may passively accept rather than critically engage with smoothed-over explanations---making them less likely to dispute an AI system's (potentially incorrect) decision~\cite{he2025conversational}. 

If we cannot rule out LLMs from producing unfaithful or placebic explanations that might unjustifiably engender more user trust~\cite{eiband2019impact}, then perhaps the XAI community should think about roles for LLMs more akin to a helpful adversary. In domains like law and economics, helpful adversaries in the form of opposing litigators and activist investors can surface the truth toward ends like justice and fair market prices. In fact, Chiang et al.~recently experimented with creating a LLM-powered \textit{devil's advocate} to assist in group decision-making regarding AI recommendations, and found that it was perceived as a useful collaborator in helping the group develop appropriate trust in another AI's recommendations~\cite{chiang2024enhancing}. We aim to apply this idea to XAI, where the devil's advocate can help users arrive at more correct interpretations of explainer outputs rather than only producing a compliant natural-language translation. In this setting, the devil's advocate has additional \textit{context} about the AI-under-test beyond its prediction/recommendation in the form of the explainer outputs, but it is possible that this context leads to multiple (perhaps conflicting) rationales about the model's behavior. Users who reckon with the incompleteness or ambiguity in an explanation should have better-justified system reliance and improved decision outcomes compared to trusting a single, plausible narrative. If we expect the latter to be the typical response when prompting an LLM to \textit{``Help me make sense...''}, then what prompting strategies could help move users beyond this?

\smallbreak
\noindent
\textbf{Prompting Approaches.} 
We think there is potential in reshaping LLM prompts as a means of producing more engaging interpretations of XAI outputs. First, it may be important to reframe the relationship between the user and the LLM in the prompt. For example, some LLM-based systems begin with an initial prompt to explain the overarching task (e.g., ``\textit{You are a helpful XAI assistant...}''). Alternatively, level-setting for more scrutiny of the XAI could work to surface alternative interpretations, e.g., ``\textit{You are a helpful XAI adversary, promoting healthy skepticism in AI...}''
Then, examples of desired inputs and outputs in the prompt can help the LLM perform better at the task (few-shot in-context learning) without fine-tuning~\cite{fewShotPaper}. 
In Table~\ref{tab:prompts}, we provide a list of prompt exemplars (organized by different strategies) that can be used as a springboard for future research and exploration. We note that these were synthesized based on shortcomings found in naively-prompted LLM translations of XAI explanations, but empirical evaluations are ultimately needed to assess the effectiveness of these strategies.

\begin{table}[t]
\centering
\small
\renewcommand{\arraystretch}{1.5}
\sffamily
\resizebox{.99\linewidth}{!}{%
\begin{tabular}{p{0.2\linewidth} p{0.8\linewidth}}
\toprule
\textbf{Adversarial Strategy}           & \textbf{Input Prompt}   \\
\midrule 
Uncertainty Awareness       & Instead of presenting every AI decision as fact, you should surface uncertainty and highlight the conditions under which an explanation might be unreliable, e.g., \textit{``The model is 78\% confident in this classification, but the decision boundary is unstable. Here are alternative inputs to probe the model...''} \\\midrule 
Alternative Explanations    & Instead of treating a single explanation as absolute, you should present multiple interpretations and counter-explanations, prompting users to compare them, e.g., \textit{``LIME suggests Feature X is most important, but SHAP highlights Feature Y instead. Here are possible reasons why they differ...''}                        \\\midrule 
Bias Detection              & Instead of smoothing potential flaws in the model, you should interrogate its outputs for potential bias or fairness issues and alert users when disparities exist, e.g., \textit{``This model’s predictions for Group A differ significantly from Group B. Here is an analysis for whether this is a fairness issue...''}                 \\\midrule 
Counterfactual Thinking     & Instead of promoting a linear exploration, you should encourage counterfactual investigation, helping users see how slight input changes would alter predictions, e.g., \textit{``If Feature X were reduced by 10\%, the model’s decision would flip. Here are some other similar scenarios that flip the model's output...''}             \\\midrule 
Scrutinize Assumptions     & Instead of letting users rely on flawed mental models, you should encourage them to re-examine their own biases when they disagree with AI predictions, e.g., \textit{``You disagree with this prediction, but could there be factors influencing the model's decision that aren't obvious?''}        \\\midrule
Explanation Audit & Instead of presenting XAI outputs as infallible, you should demonstrate that explanations are simplifications and may not fully capture model behavior, e.g., \textit{``SHAP assumes feature independence, which may not hold in this case. Here is an example of its limitations in this scenario...''}                                   \\\midrule 
User-Calibrated Depth & Instead of explaining a model's prediction in the same manner every time, you should adjust the level of explanation dynamically based on how you perceive the user's expertise and preferences, e.g., \textit{``Based on your previous questions, here is a deep dive into the feature attributions for Output X...''}  \\                 
\bottomrule
\end{tabular}}
\caption{Prompting strategies for an LLM devil's advocate that leverages XAI to \textit{challenge} AI outputs.}
\label{tab:prompts}
\end{table}

\section{LLMs for XAI, Tomorrow}
\label{sec:tomorrow}
 
With advances in generative AI coming at a rapid pace, it is hard to know exactly what the goals and methods of XAI will look like on the distant horizon. It is possible that advanced generative AI models may wholly replace some of the classifiers whose outputs and XAI-generated explanations (e.g., SHAP values, saliency maps) have been translated into narratives using LLMs. New architectures that produce and use reasoning tokens (e.g., OpenAI's o1~\cite{jaech2024openai} and o3, DeepSeek's R1~\cite{guo2025deepseek}, and Gemini 2.0 Flash Thinking~\cite{flashThinking}) are likely to improve the quality of generated responses and lead to new ways of interrogating the reliability of the model's process. A chain-of-thought reasoning step is already part of some existing XAI LLM systems like Explingo~\cite{zytek2024explingo}. But as we have seen, even with highly-skilled LLMs, users still need a method (and the will) to engage interrogatively with an AI explanation: \textit{it is a human-AI interaction challenge, not just a technology hurdle.}

In the remainder of this section, we discuss workflows and research directions that could lead users to new ways of working with XAI and LLM devil's advocates to arrive at better insights and decisions.
\\ \\
\noindent\textbf{Use LLMs to Explore Different Theories of the Case}. One key use case LLMs can help in is generating counterfactual explanations for what is needed to change a prediction based on information from the explainer. Rather than simply summarizing the explanations in natural language, the LLM can provide useful insights and counterfactual scenarios to test the limits of the explanations. Fredes et al. \cite{fredes2024using} and Giorgi et al. \cite{giorgi2024natural} present methods of using LLMs to generate natural language explanations based on a set of counterfactuals for different AI systems.

We can extend these types of approaches to allow LLMs to serve as devil's advocates for counterfactual explanations. One option is to increase the temperature of LLMs~\cite{peeperkorn2024temp, renze2024effect} in generating these explanations to provide more variance in the outputted explanation. Another direction could be to challenge the weighted outputs of the counterfactual explainer, rather than taking them as ground truth. For example, when considering a loan application, a counterfactual explainer might output that if the applicant made \$5000 more in salary, they would have been approved. But an LLM can interrogate the counterfactual given and provide alternatives for the user to analyze. LLMs could also help to highlight systematic model biases---such as an overreliance on a particular feature that may be correlated with training data biases (e.g., demographic information). With the ability to understand the model internals plus nuances of the world, an LLM can provide a combination of global and local context to dispute and improve these counterfactual explanations. 
\\ \\
\noindent\textbf{Support Evidence for a Jury of Decision Makers, Not a Mechanical Judge}.%
It is frequently discussed that different types of users (e.g., model developers versus decision makers~\cite{suh2024more, suh2023metrics, suresh2021beyond}) need different types of explanations to support their workflows, such as outlined in the study by Hoffman et al.~\cite{hoffman2023explainable}. This was echoed in recent studies that found users preferred LLM narratives over SHAP and LIME diagrams~\cite{zytek2024llms, crisan2024exploring}. It is well known in the XAI community that these commonly used diagrams are confusing to non-experts---future research should address the lack of interpretable XAI visualizations, rather than relying on LLMs to translate them for the end-user. 

As an alternative approach, LLMs can be used for different persona-based prompting regimes depending on who the explanation is targeted for. For example, in a medical diagnosis XAI system, the LLM can have a highly technical persona when providing explanations to a doctor or medical student~\cite{bussone2015role}, or use layman's terms when communicating to a patient or family member. These different levels of explanatory descriptions can be based off the same XAI output but customized for the target audience. While this might simply sound like a translation problem, LLMs could also guide the thinking of the user to challenge their assumptions. For example, when communicating to a doctor, the LLM could provide a set of possible other diagnoses with associated explanations (this would be reformulating the XAI problem as search and retrieval rather than classification). Furthermore, different modalities of information in an LLM explanation can be conveyed to the user; for instance, a natural language explanation with a supporting visualization.
\\ \\
\noindent\textbf{Design for a Productive Adversarial Process}. %
\label{sec:discomfort}%
Creating more mentally demanding workflows around AI outputs has its own tradeoffs. 
Bu{\c{c}}inca et al.'s study~\cite{buccinca2021trust} found that \textit{cognitive forcing} interventions---which prime a user's deliberative thinking about an AI output---can help the user make more correct decisions in cases where the AI model is incorrect. However, users gave these interventions the lowest favorability ratings, reporting lower trust and preference for designs that most effectively reduced overreliance. 
This presents a fundamental challenge for designing devil's advocate-style explainability systems: while friction and cognitive effort could improve decision-making, they may also reduce user satisfaction and long-term adoption. 

To combat this tradeoff, developers can consider structuring engagement around success metrics, incentives, and usability improvements. For example, we could leverage concepts from game theory~\cite{tondello2018gamification, colman2016game} to make critical engagement more rewarding by having users ``defend'' their decisions against counterfactual reasoning before finalizing an action. These systems could also incorporate progressive difficulty by starting with simple explanations and gradually introducing more complex challenges as users demonstrate comfort. Finally, creating structured workflows that acknowledge thoughtful decision-making---such as visualizing the impact of users' skepticism over time or providing trust calibration feedback---could help the user feel they are being guided through nuanced exploration of an AI system rather than interrogated for their prior beliefs. Regardless, there will inevitable difficulties in employing a devil's advocate when users are accustomed to helpful chat assistants~\cite{zhang2024sa}; research will need to explore the balance between the two personas.   


\section{Conclusion}

The widespread use of LLMs to translate XAI outputs (e.g., SHAP, LIME, CFEs) into narrative explanations assumes that natural language improves user understanding. However, recent studies suggest that this translation-based approach is insufficient and can even be counterproductive---leading to overreliance, misplaced trust, and superficial engagement with AI decisions. LLM-generated explanations often make model outputs appear more plausible rather than more transparent, and conversational XAI assistants smooth over inconsistencies which risk reinforcing cognitive biases.
To address these challenges, this workshop paper proposes a shift from LLMs as passive translators to instead serve as devil's advocates. 
Although this approach may introduce user discomfort, adversarial reasoning can begin to combat overconfidence and promote deeper engagement with AI systems. Future work should investigate how LLM-driven XAI systems can dynamically adjust explanation depth, provide active counterfactual reasoning, and challenge overreliance to ensure that explanations genuinely support better decision-making, rather than making AI outputs easier to accept.

\begin{acks}
We thank the organizers of the HCXAI workshop as well as the reviewers for their time and helpful feedback in improving the quality of our paper.

\small
\smallbreak 
\noindent 
DISTRIBUTION STATEMENT A. Approved for public release. Distribution is unlimited. This material is based upon work supported by the Combatant Commands under Air Force Contract No. FA8702-15-D-0001. Any opinions, findings, conclusions or recommendations expressed in this material are those of the author(s) and do not necessarily reflect the views of the Combatant Commands. © 2024 Massachusetts Institute of Technology. Delivered to the U.S. Government with Unlimited Rights, as defined in DFARS Part 252.227-7013 or 7014 (Feb 2014). Notwithstanding any copyright notice, U.S. Government rights in this work are defined by DFARS 252.227-7013 or DFARS 252.227-7014 as detailed above. Use of this work other than as specifically authorized by the U.S. Government may violate any copyrights that exist in this work.
\normalsize
\end{acks}

\bibliographystyle{CHI-ACM-contents/ACM-Reference-Format}
\bibliography{kgs_llms}

\appendix

\end{document}